\documentclass[aps,prd,preprintnumbers,nofootinbib,
superscriptaddress]{revtex4}
\usepackage[dvipdfmx]{graphicx}
\usepackage{bm,latexsym,amsmath,amssymb,amsfonts}
\usepackage{placeins}
\DeclareMathAlphabet{\mathpzc}{OT1}{pzc}{m}{it}
\usepackage{color}
\usepackage[normalem]{ulem}
\usepackage{hyperref}
\usepackage{subcaption}
\usepackage{braket}
\usepackage{float}

\begin{document}
\title{Renormalization group improved black holes in non-commutative momentum-dependent spacetime geometry} 

\author{Gema Ilham Baskara Darman}
\email{gema.ilham@alumni.ui.ac.id}
\author{Ar Rohim}
\email{ar.rohim@ui.ac.id}
\author{Anto Sulaksono}
\email{anto.sulaksono@sci.ui.ac.id}
\affiliation{Departemen Fisika, FMIPA, Universitas Indonesia, Depok 16424, Indonesia}

\begin{abstract}
We investigate black holes (BHs) in momentum-dependent spacetime geometry and assess its quantum Reissner-Nordstr\"om (RN) consistency with the weak gravity conjecture (WGC). Quantum corrections are introduced through the non-commutative momentum space algebra as the quantization process, and spacetime renormalization approach as a map between momentum and spacetime spaces. For the Schwarzschild case, thermodynamic analysis indicates the existence of a hot (non-zero temperature) BH remnant when evaporation stops (the entropy becomes zero), obeying a complementary third law for black hole thermodynamics.  We extend this framework to the RN solution and examine its extremal limit. For large BHs with $M \gg M_P$ ($M_P$ for Planck mass), the quantum-improved RN geometry exhibits a non-zero Hawking temperature in the extremal case, consistent with the WGC, which stipulates that extremal states should not be exactly stable or cold. The resulting momentum-dependent metric and thermodynamic properties are shown to reproduce the results derived from the Poincaré algebra (classical model) in the infrared (IR) regime.
 
\end{abstract}

\maketitle
\section{Introduction}
BHs remain among the main avenues for quantum gravity phenomenology. An approach from doubly special relativity (DSR) produced corrections to classical BH thermodynamics, leaving a remnant~\cite{Ling:2005bq, Ling:2005bp, Adler:2001vs}. This finding motivated different viewpoints on quantum-improved BHs which are inspired by gravity's rainbow and the generalized uncertainty principle (GUP)~\cite{Ling:2005bp, Galan:2006by, Myung:2006qr, Liu:2007fk, Salesi:2009kd, Gim:2014ira, Nozari:2015rza, Mu:2015qna, Kim:2016qtp, Lobo:2020oqb, Song:2025qpo, Rohim:2025gxo}. However, we offer elucidations of these approaches using the principle of relative locality~\cite{Amelino-Camelia:2011lvm} and the geometry of curved spacetime from non-commutative momentum space~\cite{Pfeifer:2021tas}. The renormalized spacetime approach~\cite{Bonanno:2000ep} utilized as a map between the momentum and spacetime spaces could also be seen as a novelty that unites the agnostic renormalization approach with the non-commutative momentum-dependent spacetime space. Additionally, we examine the theory's consistency with the WGC, a swampland conjecture~\cite{Vafa:2005ui, Brennan:2017rbf, Palti:2019pca,Loges:2019jzs}.

As reviewed in~\cite{Amelino-Camelia:2008aez,Lukierski:1992dt}, DSR—a theory proposing that the speed of light and the Planck length are both invariant scales— can emerge from the deformation of the  Poincar\'e group, which is correlated with the modification of the standard symmetries of spacetime. This deformation is based on the rainbow (energy-dependent) spacetime, meaning that the geometry of spacetime varies with the energy of the particles involved. On the other hand, quantum field theory in curved spacetime might break down before the Planck length according to string theory~\cite{Witten:2024upt}. If there are two identical particles, each with Planck mass, collide with each other, we cannot understand what happen inside the possible black hole made (at a very short distance $R=2M_P$). These points, along with other arguments explained later, motivate the consideration of a non-commutative momentum-dependent spacetime with the Planck length as a limit of the theory.

Non-commutative geometry is intuitively known as ``geometry without point" \cite{Chen:2014jwq}, thus we rely on the symmetry and the commutation relations to obtain the spacetime geometry. Utilizing non-commutative algebra could be seen as our method for quantization. On the other hand, renormalization group is an agnostic model that is the most minimal modification of a field theory that does not have a specific underlying fundamental theory \cite{Ishibashi:2021kmf,Bonanno:2000ep}. Thus, in this work, we show that the approach of renormalization group combined with non-commutative geometry can be used to explain quantum-improved black holes.

The presence of Planck scales as a limit implies the need for a deformed composition law for momentum \cite{Carmona:2019fwf, Lukierski:2002fd}. It consequently modifies the geometry of the momentum and spacetime spaces, as well as their corresponding symmetry algebras. The generators of the groups follow a non-commutative algebra formalism, which, under some parameter adjustments at the IR limit, returns to the usual Poincar\'e algebra \cite{Pfeifer:2021tas}. The deformed metric in this note has previously been derived in \cite{Pfeifer:2021tas}, which is a curved spacetime generalization of the models in \cite{Relancio:2020zok, Carmona:2021gbg}. However, our paper differs in the connection between momentum and position, where we adopt the renormalized spacetime approach and how it relates to the momentum of the test particle as a means of perceiving the momentum-dependent geometry as a quantum improvement \cite{Bonanno:2000ep, Ishibashi:2021kmf, Pawlowski:2018swz}. The focus of this paper is on the signatures of the corresponding BH spacetime, rather than the formalism of the algebra, as explained in the referenced literature.

One implication of the principle of relative locality is that the geometry of spacetime could be relative to the energy-momentum of the test particle, $k_{\mu}~$\cite{Amelino-Camelia:2011lvm}. This principle sets the basis for the momentum-dependent spacetime metric. However, a new problem arises: to study thermodynamics, we need to work with a pure spacetime-dependent metric. The transformation of the metric to a pure spacetime-dependent metric,
\begin{equation}
    g(x,k) \rightarrow g(x),
\end{equation}
on a force-free condition for the test particle, can be handled with the help of the modified dispersion relation and the scale (cutoff) identification from the renormalization group as a map between momentum and spacetime \cite{Bonanno:2000ep}. The expression for the connection between momentum and spacetime, and the corresponding dispersion relation in this work are not the same as in \cite{Pfeifer:2021tas}, in which, in that article, the selected connection is the non-linear one and the momentum and spacetime spaces are defined as cotangent spaces. 

The dispersion relation can also be defined as the relation between the mass of a test particle as the invariance in momentum space and the inner product of momenta as follows (Append. \ref{Appendix : DDR}),
\begin{equation}\label{Eq:dispersion-relation}
    m^2=g^{\mu \nu}(x,k)k_{\mu}k_{\nu}.
\end{equation}
However, if we are not looking at the momentum and spacetime spaces as cotangent bundle or without a specifically defined momentum and spacetime map, the energy $\omega$ and momentum $\vec{k}$ will be unbounded. Thus, we employ the renormalization group approach which utilize the scale (cutoff) identification between the UV and IR region and proper distance for a quantum-improved spacetime as follows \cite{Bonanno:2000ep, Ishibashi:2021kmf},
\begin{equation}\label{Eq:connection}
    k_r(r)=\frac{\xi}{d(r)},
\end{equation}
where $d(r)$ is the radial proper distance and $\xi$ is some constant, which could be used for Schwarzschild BH \cite{Bonanno:2000ep}. Whereas, for an RN BH, the proper definition of scale identification follows this equation \cite{Ishibashi:2021kmf, Pawlowski:2018swz},
\begin{equation}
    k_r^4(r)=\chi^4K(r),
\end{equation}
where $K(r)$ is the Kretschmann scalar and $\chi$ is some constant, related to the curvature of the corresponding BH. The proper distance is used for the Schwarzchild spacetime since the Ricci tensor vanishes, which is not the case for an RN BH \cite{Ishibashi:2021kmf}. Both the proper distance and the Kretschmann scalar could be utilized because they are diffeomorphism-invariant quantities. 

In 1938, Max Born proposed momentum space as a framework to unify quantum theory and relativity and introduced the concept of a reciprocity relation between momentum and spacetime \cite{Born:1938zve}.  Spacetime and momentum can be connected by a reciprocity relation between momentum and proper distance in spherically symmetric geometry (\ref{Eq:connection}), as rigorously applied in \cite{Bonanno:2000ep, Ishibashi:2021kmf}. These considerations indicate that the deformed spacetime metric $g_{\mu \nu}(x,k)$ is dual to the momentum space metric $g^{\mu \nu}(x,k)$. This approach allows us to determine the momentum space metric and thereby obtain the corresponding spacetime metric.

The model in this paper produced a remnant with zero entropy and zero heat capacity, but non-zero temperature. This possibility is discussed with its relation to the complementary third law for BH thermodynamics in \cite{Yao:2018ceg}. The condition of zero heat capacity suggests that the BH has no thermodynamical interaction with its environment (thermodynamically inert like an elementary particle) with zero entropy (leaving a stable or metastable remnant). For a comprehensive discussion on remnant, see \cite{Chen:2014jwq}. In addition to its role in preventing complete evaporation, a study of BH thermodynamics plays a non-trivial role for the swampland program \cite{Vafa:2005ui,Brennan:2017rbf,Palti:2019pca}. An effective theory of quantum black holes could be assessed thermodynamically by investigating the charged BHs and its consistency with the Weak Gravity Conjecture (WGC) \cite{Loges:2019jzs}, where quantum gravity theory should allow a state in which an extremal BH can have non-zero Hawking temperature. 

The Hawking temperature, which is one of the cornerstones of BH thermodynamics, could be approached from the surface gravity. However, it could also be evaluated from the photon energy near the horizon~\cite{Ling:2005bp,Ling:2005bq,Adler:2001vs}. In this work, we present that the photon energy approach matches the result from surface gravity specifically in the case of large BHs, which also allows us to evaluate the Hawking temperature of the large RN BH ($M^2\gg M_P^2$). We examine the resulting Hawking temperature and find its consistency with the WGC in the case of large RN BH, where we can use the photon energy approach.

Additionally, the theory in \cite{Pfeifer:2021tas}, which is used extensively in this work, introduces its own constants and parameters. However, their relations to natural constants remain undiscussed. We identify constraints on the relevant parameters based on the consequences of the resulting modified dispersion relation and thermodynamics, which are related to the Planck constants. 

The rest of the structure of this work is as follows. In Sec.~\ref{sec: metric}, we briefly review the formalism of the metric arises from the non-commutative geometry in \cite{Pfeifer:2021tas}. In Sec.~\ref{QI-BH}, we calculate the deformed metric following the non-commutative momentum-dependent geometry. We also derived the modified dispersion relation and the thermodynamics from the classical, surface gravity, and the photon energy approaches. Additionally, some relevant constants are also determined. In Sec.~\ref{sec: RN}, we discuss the solution of quantum-improved RN BH and assess its consistency with the WGC as a swampland conjecture. Sec.~\ref{sec: conclusion} is devoted to conclusion.

\section{Metric}\label{sec: metric}

The formalism presented in this section is a brief review of that presented in \cite{Pfeifer:2021tas}. This is needed to clarify the symmetry algebra of our momentum space and its relation to the corresponding spacetime. The approach we use to accommodate the Planck length is to employ the deformed composition law \cite{Amelino-Camelia:2011lvm} in a maximally symmetric momentum space for consistent physics. The resulting algebra is a noncommutative algebra rigorously derived in \cite{Pfeifer:2021tas}.

The action of the translations, boosts, and rotations generators on the momentum can be written as follows,
\begin{equation}
    k_\mu \rightarrow \tilde{k}_{\mu}=k_{\mu}+q_{\mu}T(k)+\Omega_{\rho\sigma}J^{\rho \sigma}_{\: \: \: \mu} (k),
\end{equation}
where $k_{\mu}$ is momentum, $\tilde k_{\mu}$ is the momentum after transformation, $q$ is the translation parameter, $T(k)$ is the Killing vector field for translation, $\Omega$ are the anti-symmetric rotation and boost parameters ($\Omega_{\mu \nu}=-\Omega_{\nu \mu}$), and $J^{\rho\sigma}_{\: \: \mu}$ is the Killing vector field for rotation. The maximally symmetric momentum space requires 10 symmetry generators (4 translations and 6 boosts and rotations). The representations of translations $\mathcal{T}^{\lambda}$, and boosts and rotations $\mathcal{J}^{\mu \nu}$ generators in the momentum space, respectively, are
\begin{equation}\label{eq: translations gen}
    \mathcal{T}^{\lambda}=T(k) \frac{\partial}{\partial k_{\lambda}}=\sqrt{1-K\eta^{\mu \nu}k_{\mu}k_{\nu}} \frac{\partial}{\partial k_{\lambda}},
\end{equation}
\begin{equation}\label{eq: boosts gen}
    \mathcal{J}^{\mu \nu}=J^{\mu \nu}_{\: \: \: \lambda}(k)\frac{\partial}{\partial k_{\lambda}}=k_{\rho}(\delta^{\mu}_{\: \lambda}\eta^{\nu\rho}-\delta^{\nu}_{\: \lambda}\eta^{\mu \rho})\frac{\partial}{\partial k_{\lambda}}.
\end{equation}
One form of the metric that is invariant under the transformations by the above generators on flat spacetime is the following metric,
\begin{equation}\label{eq:metric-on-flat-space}
    \zeta^{\mu \nu}(x,k)=\eta^{\mu \nu}(x)+\frac{K}{1-K\eta^{\rho \sigma}(x)k_{\sigma}k_{\rho}}\eta^{\mu \lambda}(x)k_{\lambda}\eta^{\nu \iota}(x)k_{\iota},
\end{equation}
where $\eta$ is the Minkowski metric, and $K=\pm 1/\Lambda^2$ where $\Lambda$ is in the order of Planck energy. For $K=0$, it goes back to the usual Poincar\'e algebra. A metric similar to the above metric has also been introduced in \cite{Carmona:2021gbg}.

The generators of the isometries in (\ref{eq: translations gen}) and (\ref{eq: boosts gen}) form some kind of non-commutative geometry in our maximally symmetric momentum space. The algebra is as follows,
\begin{equation}
[\mathcal{T}^{\alpha},\mathcal{T}^{\beta}]=K\mathcal{J}^{\alpha \beta},
\end{equation}
\begin{equation}  [\mathcal{T}^{\alpha},\mathcal{J}^{\beta \gamma}]=\eta^{\alpha \beta}\mathcal{T}^{\gamma}-\eta^{\alpha \gamma}\mathcal{T}^{\beta},
\end{equation}
\begin{equation}
[\mathcal{J}^{\alpha \beta},\mathcal{J}^{\gamma \delta}]=\eta^{\beta \gamma} \mathcal{J}^{\alpha \delta}-\eta^{\alpha \delta} \mathcal{J}^{\beta \gamma}-\eta^{\beta \gamma} \mathcal{J}^{\alpha \delta}+\eta^{\alpha \delta} \mathcal{J}^{\beta \gamma},
\end{equation}
where it returns to the Poincar\'e algebra or commutative geometry on the maximally symmetric momentum space at $K=0$. We will use the metric for flat spacetime in (\ref{eq:metric-on-flat-space}) to be generalized to the curved spacetime in a consistent way.

We can lift the flat space metric (\ref{eq:metric-on-flat-space}) to curved space by using the tetrad of the metric via a conjecture in \cite{Pfeifer:2021tas}, such that
\begin{equation}\label{eq: curved space metric}
    g^{\mu \nu}(x,k)=a^{\mu \nu}(x)+\frac{K}{1-Ka^{\lambda \sigma}(x)k_{\lambda}k_{\sigma}}a^{\mu \alpha}(x)k_{\alpha}a^{\nu \beta}(x)k_{\beta},
\end{equation}
where the curved spacetime metric (before deformation) can be defined as $a^{\mu \nu}(x) = e^{\mu}_{\: \alpha}(x) \eta^{\alpha \beta} e^{\nu}_{\: \beta}(x)$, with $e^{\mu}_{\: \alpha}(x)$ denoting the vielbein, and $\eta^{\alpha \beta}$ representing the Minkowski metric. The above metric can be used to construct any classical metric $a^{\mu \nu}(x)$ to be a momentum-dependent one.

It should be emphasized that $k$ is the 4-momentum of the test particle in the corresponding spacetime metric. An implied physical argument is that different test particles with different masses experience spacetime differently in a covariant manner, following the principle of relative locality \cite{Amelino-Camelia:2011lvm}. The corresponding metric on spacetime $g_{\mu \nu}(x,k)$ is related to the momentum space metric $g^{\mu \nu}(x,k)$ such that one serves as the inverse of the other in the given context. In Sec. \ref{QI-BH}, we will use the momentum space metric $g^{\mu \nu}(x,k)$ to find the dispersion relation, which will be relevant to study the metric and relevant thermodynamic properties.

For convenience, we can study the condition where $k^2 = a^{\rho \sigma}(x)k_{\rho}k_{\sigma} = 0$, with $k_{\mu}$ as the 4-momentum and $a^{\rho \sigma}(x)$ as defined above; this could be interpreted as the massless test particle condition before deformation. Hence, one can write the metric as follows,
\begin{equation}\label{eq: improved-metric}
    g^{\mu \nu}(x,k)=a^{\mu \nu}(x)+Ka^{\mu\alpha}(x)k_{\alpha}a^{\nu \beta}(x)k_{\beta},
\end{equation}
where the momentum will be employed as the cutoff (scale) identification. This metric has been derived in \cite{Pfeifer:2021tas}, which we will utilize extensively in this work.

\section{Quantum-improved Schwarzschild BH}\label{QI-BH}

In this section, we investigate the quantum-improved Schwarzschild metric, which can be generalized to other spherically symmetric BHs. The static spherically symmetric metric of the Schwarzschild BH can be written as follows,
\begin{equation}
    ds^2=a_{\mu \nu}(x)dx^{\mu}dx^{\nu}=f(r)dt^2-\frac{dr^2}{f(r)}-r^2(d\theta^2+\sin^2 \theta d\phi^2),
\end{equation}
where $a_{\mu\nu}(x)$ is the classical curved spacetime metric, $f(r)=1-\frac{2GM}{r}$ is the usual lapse function, and $M$ is the mass of the BH. 

The deformation on the classical Schwarzschild metric in momentum space, following (\ref{eq: improved-metric}), is as follows,
\begin{align}\label{eq: metric}
    g^{tt}(r,\omega) &= \frac{1}{f(r)}\left(1+\frac{K\omega^2}{f(r)}\right), \\
    g^{rr}(r,k_r) &= -f(r)(1-Kf(r)k_r^2), \\
    g^{\theta \theta}(r)&=-\frac{1}{r^2},\\
    g^{\phi \phi} (r,\theta)&=-\frac{1}{r^2 \sin^2 \theta}, 
\end{align}
assuming $k=(\omega,-k_r,-k_{\theta}=0,-k_{\phi}=0)$ for the radial free fall condition on the test particle, in which, $\omega$ is the energy and $k_r$ is the radial momentum of the test particle in natural units. This assumption leaves the angular components of the metric to be non-deformed. A similar metric has also been derived in \cite{Relancio:2020zok}. 

\subsection{Modified dispersion relation and other parameters}\label{sec: MDR and constants}

The dispersion relation describes how the test particle's energy depends on momentum, and mass as the invariance under Lorentz transformations. In Minkowski space, it is given by $m^2=\eta^{\mu \nu}(x)k_{\mu}k_{\nu}$. For momentum-dependent geometries and the deformed flat spacetime metric in (\ref{eq:metric-on-flat-space}), it becomes $m^2=\zeta^{\mu \nu}(x,k)k_{\mu}k_{\nu}$. Using tetrad and the metric from (\ref{eq: curved space metric}), the relation is $m^2=g^{\mu\nu}(x,k)k_{\mu}k_{\nu}$. In \cite{Pfeifer:2021tas}, the dispersion relation is defined from the relation between momentum and spacetime with some non-linear connection, albeit it stems from the same principle as in this work that the dispersion relation contains Lorentz invariance or Casimir operator. The Lorentz invariability means that these relations are unchanged by Lorentz transformations (see Appendix (\ref{Appendix : DDR})). Based on these considerations, the dispersion relation of the test particle is as follows,
\begin{equation}
    m^2=\frac{\omega^2}{f(r)}\left(1+\frac{K\omega^2}{f(r)}\right)-k_r^2f(r)(1-Kf(r)k_r^2).
\end{equation}
The energy of the test particle, neglecting a small $K^2$ term, with mass $m$ reads as follows,
\begin{equation}\label{Eq: energy}
    \frac{\omega^2}{f(r)}=\frac{1}{2K}(\sqrt{1+4K(k_r^2f(r)+m^2)}-1),
\end{equation}
which, in addition to (\ref{Eq:connection}), helps evaluate the pure spacetime metric $g_{\mu \nu}(x)$.

For a well-defined energy and momentum, due to the presence of singularities in $f(r)=1-2GM/r$, we can define the scaled energy $\tilde \omega(r)$ and the scaled radial momentum $\tilde k_r(r)$ as follows,
\begin{equation}\label{eq: scaled energy}
    \tilde \omega^2 (r)= \frac{\omega^2}{f(r)},
\end{equation}
\begin{equation}\label{eq: scaled-kr2}
    \tilde k_r^2 (r)= k_r^2 f(r),
\end{equation}
which then, as a quantum-improved model, we extend the definition of radial momentum \cite{Bonanno:2000ep, Ishibashi:2021kmf} to the scaled radial momentum $\tilde k_r(r)$ as follows,
\begin{equation}\label{eq: scaled-kr-cutoff}
    \tilde k_r (r)= \frac{\xi}{d(r)}.
\end{equation}
Here, $\xi$ is a constant to be determined, and $d(r)$ is the radial proper distance. We define the scaled radial momentum as the cutoff identification. The results give singularity-free scaled radial momentum $\tilde k_r(r)$ and scaled energy $\tilde \omega (r)$. The scaled energy and momentum result agrees with the results in \cite{Ling:2005bq, Ling:2005bp}.

The scaled energy $ \tilde \omega^2$ and scaled momentum $ \tilde k_r^2$ discussed above can be traced from the inner product of 4-momentum as follows,
\begin{equation}
    k \cdot k=a^{\mu \nu}(x)k_\mu k_\nu=\frac{\omega^2}{f(r)}-k_r^2f(r)=\tilde \omega(r)^2-\tilde k_r(r)^2,  
\end{equation}
in which we can state that the scaled energy is formally $\tilde \omega^2=a^{tt}(k_t)^2$ and the scaled radial momentum reads $\tilde k_r(r)^2=a^{rr}(k_r)^2$, for 4-momentum vector $k=(\omega,-k_r,0,0)$ with $a^{\mu \nu}$ as the classical metric. Since both scaled energy and scaled radial momentum are scalars, they can be used as scale identifications, which are connected to other scalars e.g. proper distance and Kretschmann scalar. 

From the above calculations and correlation with the results from DSR in \cite{Ling:2005bq,Ling:2005bp}, the scaled energy reads
\begin{equation}\label{eq: scaled energy 2}
    \tilde \omega (r)= \left[\frac{M_P^2}{2}\left(1-\sqrt{1-\frac{4}{M_P^2}(\tilde k_r^2+m^2)} \right)\right]^{1/2},
\end{equation}
where it also implies,
\begin{equation}
    K=-\frac{1}{\Lambda^2}=-\frac{1}{M_P^2}=-\frac{1}{E_P^2},
\end{equation}
in which $\Lambda = M_P=E_P$ for $E_P$ as the Planck energy, in natural units ($\hbar=c=1$). The case $K\geq0$, from the square-root sign in Eq. (\ref{Eq: energy}), leads to complete evaporation.

In \cite{Bonanno:2000ep, Binetti:2022xdi}, with some parameterization, the proper distance reads $d(r\rightarrow2GM)=\pi GM$ on the classical Schwarzschild horizon. At the same time, imposing $g_{tt}=0$ in (\ref{eq: metric}), the horizon of the BH in this paper reads $r_H=2GM$. Thus, the scaled energy at the horizon (for massless test particle) reads
\begin{equation}
    \tilde \omega(r\rightarrow r_H)=\left[\frac{M_P^2}{2}\left(1-\sqrt{1-\frac{4 \xi^2}{\pi^2 G^2M_P^2M^2}} \right)\right]^{1/2},
\end{equation}
which, with an appropriate parameter choice, implies $\xi=\pi GM_P^2$. Using $G=1/(8\pi M_P^2)$ and natural units ($\hbar =c=1$), this gives $\xi =1/8$, so that the Hawking temperature reproduces the classical limit $T_H=\tilde \omega_{ph}(r_H)=1/(8\pi GM)$ \cite{Ling:2005bp,Ling:2005bq}.

The square root sign in (\ref{Eq: energy}) gives a constraint on each parameter involved only for $K<0$. Thus, the test particle's mass $m$ is bounded from the above as follows,
\begin{equation}
    m \leq \frac{M_P}{2},
\end{equation}
consistent with \cite{Ling:2005bq}, since $\Lambda = M_P$ in which $M_P$ is the Planck mass. Note that, since $c=1$, Planck mass and energy have the same dimension.

On the other hand, the square root sign in (\ref{Eq: energy}) also gives a constraint to the radial momentum as follows,
\begin{equation}\label{eq: radial scaled momentum}
    \tilde k_r^2\leq \frac{M_P^2}{4}-m^2,
\end{equation}
which corresponds to DSR condition in \cite{Ling:2005bq}. Consequently, the radial proper distance $d(r)$ will also be constrained as follows,
\begin{equation}\label{eq: classical distance}
    d(r)^2 \geq \frac{4\pi^2 G^2M_P^4}{M_P^2-4m^2},
\end{equation}
where condition $m > M_P/2$ yields complex $d(r)$ and $\tilde k_r(r)$. Eq.(\ref{eq: classical distance}) also imply that $M\geq 2M_P$. From the value of proper distance, we can see that the theory does not have UV completion on the first-order improvement. The UV completion could naturally arise from higher-degree iterations on improvements of the classical metric with Eq.(\ref{eq: improved-metric}).

\subsection{Thermodynamics}\label{sec: thermodynamics}

The study of BH thermodynamics starts from its Hawking temperature $T_H$. The Hawking temperature can be obtained as follows,
\begin{equation}\label{eq: temperature from surface gravity}
    T_H=\frac{1}{4\pi}\lim_{r \rightarrow r_H=2GM}\sqrt{-g^{tt}g^{rr}} \left(\frac{\partial g_{tt}}{\partial r}\right),
\end{equation}
which is the formalism following the surface gravity approach.

From Sec.~\ref{sec: MDR and constants}, the covariant metric $g_{\mu \nu}$ reads
\begin{align}\label{Eq: covariant metric}
    g_{tt}(r,\omega) &= \frac{f(r)}{1+K\tilde \omega^2}, \\
    g_{rr}(r,k_r) &= -\frac{1}{f(r)(1-K\tilde k_r^2)}, \\
    g_{\theta \theta}(r)&=-r^2,\\
    g_{\phi \phi} (r,\theta)&=-r^2 \sin^2 \theta, 
\end{align}
where $\tilde \omega(r)^2=\omega^2/f(r)$, $\tilde k_r^2=k_r^2f(r)$, and $f(r)=1-2GM/r$. From Sec.~\ref{sec: MDR and constants}, we have obtained the expressions of the scaled energy $\tilde \omega(r)$, the radial scaled momentum $\tilde k_r (r)$, and parameter $K=-1/\Lambda^2$.

Thus, the Hawking temperature reads
\begin{equation}\label{eq: TH-lambda}
    T_H=\frac{f'(r_H)}{4\pi}\sqrt{\frac{1+(\tilde k_r^2(r_H)/\Lambda^2)}{1-(\tilde \omega^2(r_H)/\Lambda^2)}}=\frac{1}{8\pi GM}\sqrt{\frac{1+(\tilde k_r^2(r_H)/\Lambda^2)}{1-(\tilde \omega^2(r_H)/\Lambda^2)}},
\end{equation}
where for a very large $\Lambda$ ($\Lambda \rightarrow \infty$), the Hawking temperature reads $T_H=1/(8\pi GM)$. The horizon radius is $r_H=2GM$, which makes $f'(r_H)=1/(2GM)$.

Thus, the Hawking temperature becomes 
\begin{equation}
    T_H=\frac{1}{8\pi GM \sqrt{2}}\sqrt{\left(1+\frac{\Lambda^2d^2(r_H)}{\xi^2}\right)\left(1-\sqrt{1-\frac{4\xi^2}{\Lambda^2d^2(r_H)}}\right)},
\end{equation}
in which, the scaled radial momentum $\tilde k_r$ and the scaled energy $\tilde \omega$ read

\begin{equation}
    \tilde k_r (r)= \frac{\xi}{d(r)},
\end{equation}

\begin{equation}
    \tilde \omega(r)=\left[\frac{M_P^2}{2}\left(1-\sqrt{1-\frac{4\xi^2}{M_P^2d(r)^2}} \right)\right]^{1/2},
\end{equation}
where $\Lambda=M_P=E_P$ (see Sec.~\ref{sec: MDR and constants}) in natural units ($\hbar=c=1$).

Since $d(r_H)=\pi GM$ \cite{Bonanno:2000ep,Binetti:2022xdi, Ishibashi:2021kmf}, the Hawking temperature becomes
\begin{equation}
    T_H=\frac{1}{8\pi GM\sqrt{2}}\sqrt{\left(1+\frac{\pi^2M^2G^2M_P^2}{\xi^2}\right)\left(1-\sqrt{1-\frac{4\xi^2}{\pi^2M^2G^2M_P^2}} \right)},
\end{equation}
where with adjustments on some constants become
\begin{equation}
    T_H=\frac{1}{8\pi GM\sqrt{2}}\sqrt{\left(1+\frac{M^2}{M_P^2}\right)\left(1-\sqrt{1-\frac{4M_P^2}{M^2}}\right)}.
\end{equation}
It could easily be checked that it returns to $T_H=1/(8\pi GM)$ and $T_H=\tilde \omega(r_H)$ at the limit $M \gg M_P$, which is consistent with the classical Schwarzschild and the result from approaching the temperature as the photon energy \cite{Ling:2005bq,Ling:2005bp}, respectively. The surface gravity approach also presents some correction term, which will be significant on small scales ($M \sim M_P$).
\begin{figure}[h]
\includegraphics[width=0.7\textwidth]
{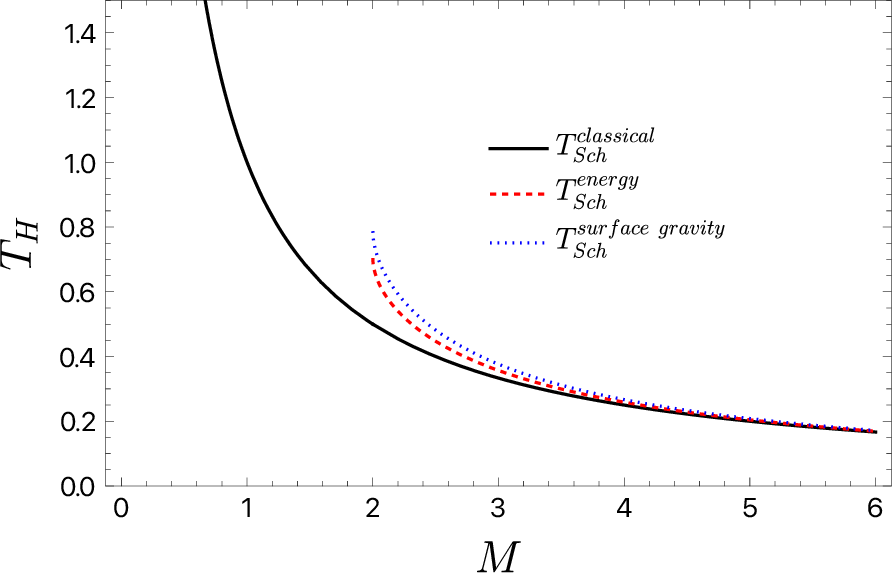}
\caption{The comparison of Hawking temperature of quantum improved BHs to that of classical one as a function of mass with $M_P=1$. The mass stops decreasing for the cases of $T_H^{energy}$ and $T_H^{surface~gravity}$ at $M=2M_P=2$.}
\label{fig: TvsM}
\end{figure}
For clarity, since $\xi=1/8=\pi GM_P^2$ (see Sec.~\ref{sec: MDR and constants}), the Hawking temperature for the classical model, the photon energy approach, and the surface gravity one read as follows, respectively,
\begin{eqnarray}
&&T_{Sch}^{classical}=\frac{1}{8\pi GM}=\frac{M_P^2}{M},
    \label{TClassical}\\
&&T_{Sch}^{energy}=\sqrt{\frac{M_P^2}{2}\left(1-\sqrt{1-\frac{4M_P^2}{M^2}} \right)},
\label{Tenergy}\\
&&T_{Sch}^{surface~gravity}=\frac{M_P^2}{M\sqrt{2}}\sqrt{\left(1+\frac{M^2}{M_P^2}\right)\left(1-\sqrt{1-\frac{4M_P^2}{M^2}} \right)},
    \label{Tsurfacegravity}
\end{eqnarray}
where $T^{surface~gravity}_{Sch}$ approach $T^{energy}_{Sch.}$ in case $M \gg M_P$ (large BH). In Fig.~\ref{fig: TvsM}, the photon energy approach and the surface gravity lead to remnant mass $M_r$ as follows,
\begin{equation}
    M_{r}=2M_P,
\end{equation}
since the BH temperature should be real. We generalize this approach to the RN BH and study the consistency with respect to WGC.

We next turn the discussion to the behavior of the heat capacity, which can be derived through the following relation
\begin{eqnarray}
C=T{\partial S\over \partial T}={\partial M\over \partial T}, 
\end{eqnarray}
where $S$ is the entropy of the BH. From the BH temperatures given in Eqs.~\eqref{TClassical}-\eqref{Tsurfacegravity}, we have the corresponding heat capacities as follows,
\begin{eqnarray}
    &&C^{classical}_{Sch.}=-{M^2\over M^2_P},\\
    && C^{energy}_{Sch.}=-{M^3\over \sqrt{2} M^3_P}\sqrt{\left(1-{4 M^2_P\over M^2}\right)\left(1-\sqrt{1-{4 M^2_P\over M^2}}\right)},\\
     && C^{surface~gravity}_{Sch.}=-{M_P^4\over M^4} {\sqrt{\left(2-{8M^2_P\over M^2}\right)\left({1+{M^2\over M^2_P}}\right)\left(1-\sqrt{1-{4M^2_P\over M^2}}\right)}\over \left(6+{M^2\over M^2_P}\left(1+\sqrt{1-{4M^2_P\over M^2}}\right)\right)}.
\end{eqnarray}
\begin{figure}[h]
\includegraphics[width=0.7\textwidth]{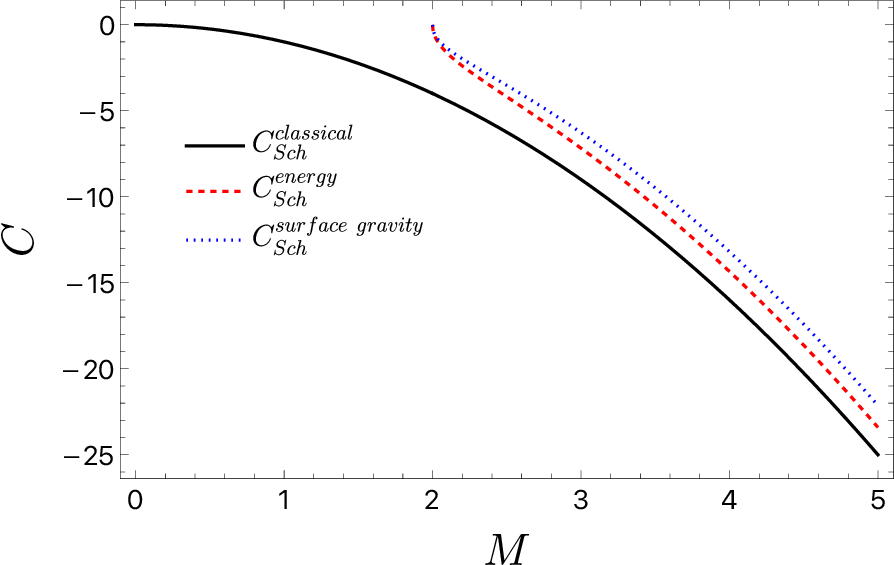}
\caption{Plot of heat capacity of the BH as a function of mass $C(M)$ with $M_P=1$.}
\label{fig:CvsM}
\end{figure}
In Fig. \ref{fig:CvsM}, we demonstrate the behavior of the heat capacity as a function of mass $M$. From this figure, one can see that the heat capacity of the classical Schwarzschild BH vanishes as the mass $M$ decreases and the BH evaporates. However, for the photon energy approach and the surface gravity method, the heat capacity vanishes as the mass $M$ approaches the remnant mass $M=2M_P$, indicating a critical point for phase transition. Since $dM= C~dT$ for Schwarzschild BH, the zero heat capacity of the remnant indicates no further evaporation at $M_r=2M_P$, while the heat capacity of classical Schwarzschild reaches zero only at $M=0$. At $M_r=2M_P$, we can set the entropy to be zero as the initial condition of the integral.

The energy loss is dominated by photons, which allows us to use the Stefan-Boltzmann law to estimate the mass and energy output as a function of time~\cite{Adler:2001vs}. The proportionality rate of energy loss, to validate the argument on the presence of BH remnant, is as follows~\cite{Cavaglia:2003qk},
\begin{equation}\label{Eq: Stefan-Boltzmann}
    \frac{dM}{dt} \propto AT^4,
\end{equation}
where $A$ denotes the area, and $T$ is the Hawking temperature. Here, we can evaluate the estimations of $dM/dt$ for each case.

\begin{figure}[h!]
\includegraphics[width=0.7\textwidth]{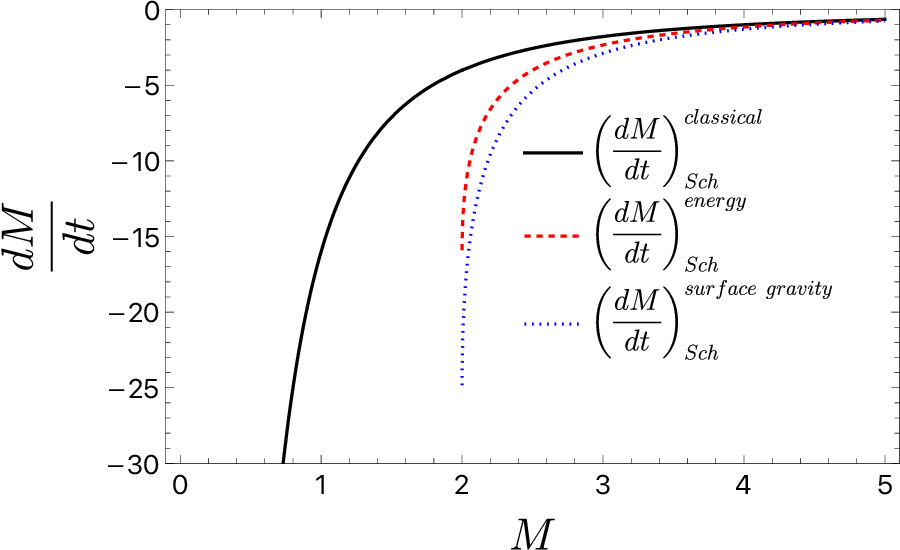}
\caption{The emission rate for the BH as a function of mass $dM/dt$ with $M_P=1$.}
\label{fig:dmdt}
\end{figure}

The emission rate for classical Schwarzschild reads~\cite{Ali:2014xqa}
\begin{equation}
    \left(\frac{dM}{dt}\right)^{classical}_{Sch}= -M_P^3 \frac{\mu}{t_P}M^{-2},
\end{equation}
where $t_P$ is the Planck time in natural units, and the form of $\mu$ can be found in \cite{Adler:2001vs, Cavaglia:2003qk, Cavaglia:2004jw}. It can be seen that the emission rate is infinite at $M\rightarrow 0$ for classical Schwarzschild. Thus, utilizing Eq.~(\ref{Eq: Stefan-Boltzmann}), we obtain
\begin{equation}
    \frac{dM}{dt}=\left(\frac{dM}{dt}\right)^{classical}_{Sch}\left(\frac{T}{T_{Sch}^{classical}}\right)^4,
\end{equation}
allowing us to calculate $dM/dt$ for Hawking temperature of each model. The results are as follows,
\begin{equation}\label{Eq: emission-energy}
    \left(\frac{dM}{dt}\right)^{energy}_{Sch}=-\frac{\mu M^2}{4M_P t_P}\left(1-\sqrt{1-\frac{4M_P^2}{M^2}}\right)^2,
\end{equation}
\begin{equation}\label{Eq: emission-surface gravity}
    \left(\frac{dM}{dt}\right)^{surface~gravity}_{Sch}=-\frac{\mu M_P^3}{4M^2t_P}\left(1+\frac{M^2}{M_P^2}\right)^2\left(1-\sqrt{1-\frac{4M_P^2}{M^2}}\right)^2,
\end{equation}
where both results suggest the presence of minimum mass $M_r=2M_P$, and do not allow the mass to be lower than $M_r=2M_P$ even though the emission rate is non zero (see Fig. \ref{fig:dmdt}).
It could also be noted that for the large mass case ($M\gg M_P$), both results are identical. At $M=2M_P$, the emission rate in Eq.~(\ref{Eq: emission-energy}) and Eq.~(\ref{Eq: emission-surface gravity}) could be negligible since $M_P$ is small, which could be called a meta-stable remnant~\cite{Chen:2014jwq}.

To emphasize the behavior of the remnant, we need to analyze the entropy of each model. The entropy reads
\begin{equation}
    S^{classical}_{Sch}=\frac{M^2}{2M_P^2},
\end{equation}
while the entropy for the energy method is logarithmic form, in which the entropy must be zero at $2M_P$ to prevent complex value as follows,
\begin{equation}
    \begin{split}
    S^{energy}_{Sch}=\frac{\sqrt{2}}{4M_P^2}[(M+M_P)\sqrt{\frac{M(M+2M_P)}{2}}-(M-M_P)\sqrt{\frac{M(M-2M_P)}{2}}\\-M_P^2(6+\frac{1}{\sqrt{2}}\ln |\frac{(M+M_P+\sqrt{M(M+2M_P)})(M-M_P+\sqrt{M(M-2M_P)})}{(3+2\sqrt{2})M_P^2}|)].
    \end{split}
\end{equation}

The entropy derived from surface gravity can only be evaluated numerically, in which we evaluate the following integration,
\begin{equation}
    S^{surface~gravity}_{Sch}=\int \frac{dM}{T^{surface~gravity}_{Sch}}=\frac{2}{M_P}\int \frac{M^{3/2}dM}{\sqrt{M_P^2+M^2}(\sqrt{M+2M_P}-\sqrt{M-2M_P})},
\end{equation}
\begin{figure}[h!]
\includegraphics[width=0.9\textwidth]{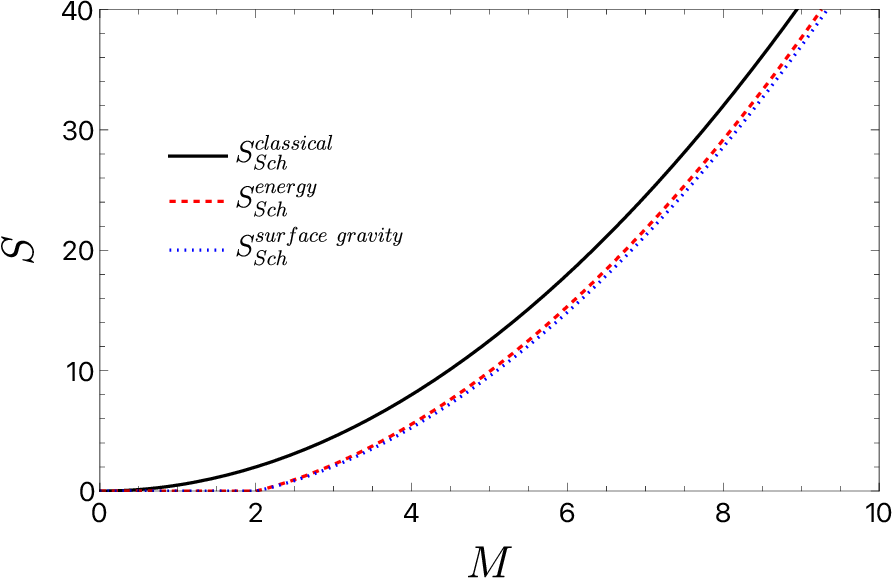}
\caption{The entropy for the BH as a function of mass $S(M)$ with $M_P=1$.}
\label{fig:SvsM}
\end{figure}
and the plot is in Fig.~\ref{fig:SvsM}. The entropy is zero at $M=2M_P$ indicating the existence of a stable remnant. The entropy of the remnant is zero while the Hawking temperature is non zero, obeying complementary third law of thermodynamics \cite{Yao:2018ceg}.

\section{Quantum-improved RN BH and WGC}\label{sec: RN}

The Hawking temperature for the RN solution can be evaluated as in Eq.(\ref{eq: temperature from surface gravity}). The differences are in the definitions of the lapse function $f(r)$ and the scale (cutoff) identification $\tilde {k_r}$. In this section, we first present the general expressions, then specialize to the extremal case $|e|=M$ with a focus on the large-mass $M \gg M_P$, which is the regime where the model in our work is consistent with the WGC.

The lapse function for the classical RN BH reads as follows,
\begin{equation}
    f_{RN}(r)=1-\frac{2M}{r}+\frac{e^2}{r^2},
\end{equation}
in which the horizons read $r_{H\pm}=M\pm\sqrt{M^2-e^2}$ where $e$ is the electric charge. In this case, we set $G=1$ so that $e$ and $M$ have the same dimensions, enabling convenient analysis of consistency with the WGC. In addition, looking at $g_{tt}$ in Eq. (\ref{Eq: covariant metric}), the horizon (from $g_{tt}=0$) is conveniently the same as the classical solution.

It can be easily obtained that the temperature vanishes for both the classical and quantum-improved extremal RN solution from the surface gravity approach, since the first derivative of $f_{RN}(r_{H,ext})$ vanishes. However, at the limit of $M\gg M_P$, the Hawking temperature from surface gravity approaches the free-falling photon energy near the horizon, which has been shown in Sec.~\ref{sec: MDR and constants}. This is also consistent with the approach in \cite{Ling:2005bp,Ling:2005bq,Adler:2001vs}. Consequently, we are able to investigate the Hawking temperature of the extremal BH with $M \gg M_P$ via its scaled photon energy $\tilde\omega_{ph}$.

With this context established, we need to define the connection between momentum and spacetime spaces which, in our work, is related to the scale identification in renormalization group. The scaled radial momentum, which becomes the scale identification, is defined as \cite{Ishibashi:2021kmf, Pawlowski:2018swz}
\begin{equation}
    \tilde k_r^4(r)=\chi^4K_{RN}(r),
\end{equation}
where the Kretschmann scalar as the scale identification reads \cite{Ishibashi:2021kmf}
\begin{equation}
    K_{RN}(r)=\frac{8}{r^8}(6M^2r^2-12Me^2r+7e^4),
\end{equation}
for classical RN BH, $\chi$ is some constant with the same dimension as the squared of mass. The Kretschmann scalar is used for scale identification because it is a diffeomorphism-invariant quantity of 4-momentum \cite{Ishibashi:2021kmf, Pawlowski:2018swz}. However, since the Ricci tensor vanishes in the Schwarzschild BH, curvature-related scale identification is prohibited. This motivates the use of the proper distance for scale identification in Schwarzschild. The definition of scaled energy still follows Eq.(\ref{eq: scaled energy}), as the lapse function $f(r)$ is absorbed by redefining the scale (cutoff) identification.

\begin{figure}[h]
\includegraphics[width=0.7\textwidth]{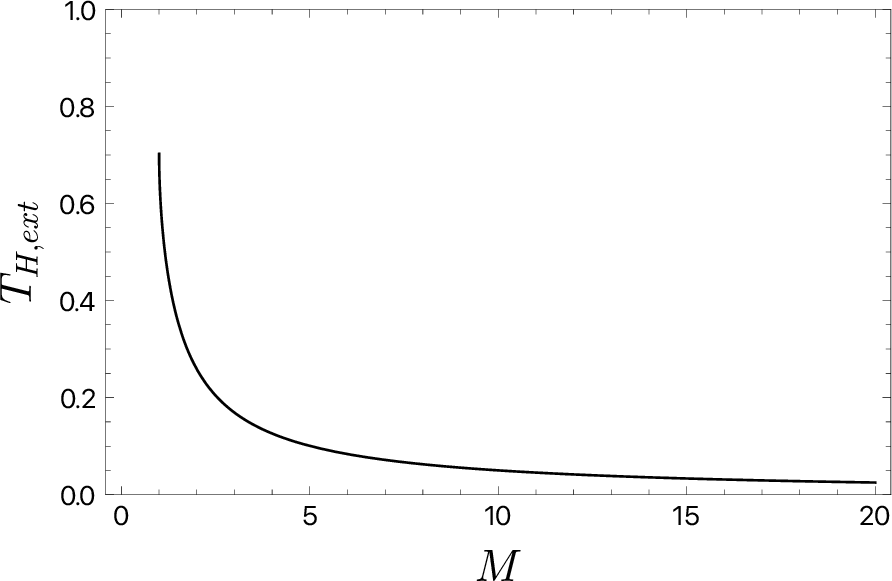}
\caption{The extremal RN Hawking temperature $T_{H,ext}(M)$ with $M_P=1$ and $\chi^2=1/(8\sqrt{2})$. $T_{H,ext}\rightarrow 0$ only at $M\rightarrow\infty$, for the energy approach.}
\label{fig: T_extvsM}
\end{figure}
In case of $M \gg M_P$, the Hawking temperature for the extremal BH $r_+=r_-=M$ reads
\begin{equation}
    T_{H,ext}=\tilde \omega_{ph} = \left[\frac{M_P^2}{2}\left(1-\sqrt{1-\frac{8 \sqrt{2}\chi^2}{M_P^2M^2}}\right)\right]^{1/2},
\end{equation}
where $K_{RN,ext}(M)=8/M^4$ and $|e|=M$. The Hawking temperature for the extremal RN solution from the photon energy approach, which is shown in Fig.~\ref{fig: T_extvsM}, does not have non-zero temperature state. This indicates that the extremal RN BH in this work is unstable, except when it reaches the remnant mass.

Consequently, the remnant mass reads
\begin{equation}
    M_{r,ext}=(128)^{1/4}\frac{\chi}{M_P}.
\end{equation}
We note that the remnant's charge-to-mass ratio $|e|/M=1$ continues to saturate the extremality bound, consistent with the WGC expectation. The theory, in the case of a large BH $M \gg M_P$ for RN BH, lifts the Hawking temperature to be nonzero, even for the extremal case, and does not allow a naked singularity. Consequently, the constant $\chi$ has to be as follows,
\begin{equation}
    \chi \gg \frac{M_P^2}{(128)^{1/4}},
\end{equation}
in order to use the photon energy approach in any regime of the black hole's mass/energy.

\section{Conclusion}\label{sec: conclusion}

We investigated a possibility that the non-commutative momentum-dependent geometry with renormalization group approach as a map between momentum and spacetime could be a novelty that technically sound. We found that the results from Sec. \ref{sec: MDR and constants} fit nicely with those from the DSR approach \cite{Ling:2005bq, Ling:2005bp} and GUP approach \cite{Chen:2014jwq}. This explains a possible underlying theory behind the minimum modification introduced by renormalization approach, which could be non-commutative momentum dependent geometry. In Sec. \ref{sec: RN}, we investigate RN BH to study the possible extension of the theory.

In the regime of large BHs, $M \gg M_P$, the theory is consistent with the WGC, which implies that the quantum-improved extremal RN BH should not be too stable (have zero temperature). We have shown in Fig.~\ref{fig: T_extvsM} that the large quantum-improved extremal RN in this paper, or from the photon energy approach, has non-zero Hawking temperature. On the other hand, both quantum-improved Schwarzschild and RN BHs leave thermodynamic remnants with zero entropy but non-zero Hawking temperature, obeying the third law of thermodynamics \cite{Yao:2018ceg}. This is consistent with the common principle that the quantum-improved BH should not completely evaporate. 

From Eq. (\ref{eq: classical distance}), we can conclude that the model has no UV completion. It is limited to some minimum proper distance and energy-momentum. However, the formal structure of the deformation suggests that the UV completion could be achieved from iteration on the quantum improvements as follows,
\begin{equation}
    a(x) \rightarrow g(x,k) \rightarrow h(x,k) \rightarrow ...,
\end{equation}
Here, $a(x)$ denotes the classical metric. Next, $g(x,k)$ represents a quantum improvement of $a(x)$, while $h(x,k)$ is a quantum improvement of $g(x,k)$. Whether Eq. (\ref{eq: improved-metric}) allows us to repeatedly lift the metric to achieve UV completion remains to be investigated.

The quantum-improved metric also yielded a modified dispersion relation, leading to a complete spacetime description of the metric along with its thermodynamics. With the help of some redefinition to remove the singularity on energy and momentum (Eq. (\ref{eq: scaled energy}) and (\ref{eq: scaled-kr2})), and relating them to diffeomorphism-invariant scalars, e.g., proper distance and Kretschmann scalar, the thermodynamics and consistencies with WGC can be studied further. 

Other BH models could be investigated using the methods employed in this paper, provided the quantum improvements are consistent with some conjectures of quantum gravity, including the WGC. Further progress requires a UV-complete extension of the model to analyze the UV region. On the other hand, we can note that the theory in this paper reduces to classical Schwarzschild and RN BHs with parameter adjustments, demonstrating consistency at the IR limit.



\begin{appendix}


\section{Modified dispersion relation as Lorentz invariance in momentum space} \label{Appendix : DDR}

The well-known dispersion relation in Minkowski spacetime, $m^2=\omega^2-\vec{k}^2$, can be obtained by the following formalism,
\begin{equation}
    m^2=\eta^{\mu \nu}k_{\mu}k_{\nu},
\end{equation}
where $\eta$ is the Minkowski metric. In DSR's case where Planck length is the limit of theory, the flat space metric in momentum space as in (\ref{eq:metric-on-flat-space}) from \cite{Pfeifer:2021tas} has been defined.

Thus, the Minkowski metric is modified as follows,
\begin{equation}
    \eta^{\mu \nu} \rightarrow \zeta^{\mu \nu},
\end{equation}
where $\zeta$ is the momentum space metric on flat spacetime. The corresponding modified dispersion relation becomes
\begin{equation}
    m^2=\zeta^{\mu \nu}k_{\mu}k_{\nu}.
\end{equation}
Generalizing the dispersion relation to the curved momentum space can be done via tetrad as follows,
\begin{equation}
    m^2=e^{\mu}_{\:\alpha}(x)\zeta^{\alpha \beta} e^{\nu}_{\:\beta}(x) e^{\alpha}_{\:\mu}(x)k_{\alpha}e^{\beta}_{\:\nu}(x)k_{\beta},
\end{equation}
where $g^{\mu \nu}=e^{\mu}_{\:\alpha}(x)\zeta^{\alpha \beta}e^{\nu}_{\:\beta}(x)$ and $\bar{k}_{\mu}=e^{\alpha}_{\:\mu}(x)k_{\alpha}$. The dummy bar index on momentum can be dropped so that the dispersion relation becomes as follows,
\begin{equation}
    m^2=g^{\mu\nu}(x,k)k_{\mu}k_{\nu},
\end{equation}
where the metric $g^{\mu \nu}(x,k)$ is the curved momentum space metric as in (\ref{eq: curved space metric}).

If one performs Lorentz transformation to the momentum,
\begin{equation}
    k_{\mu} \rightarrow k'_{\mu}=\Lambda^{\alpha}_{\:\mu}k_{\alpha},
\end{equation}
the dispersion relation is proved to be invariant as follows,
\begin{equation}
    m^2=g^{\mu\nu}(x,k)k'_{\mu}k'_{\nu}=\Lambda_{\:\alpha}^{\mu}\Lambda_{\:\beta}^{\nu}g^{\alpha \beta}(x,k)k_{\mu}k_{\nu}=g^{\mu\nu}(x,k)k_{\mu}k_{\nu},
\end{equation}
where $\Lambda_{\:\alpha}^{\mu}\Lambda_{\:\beta}^{\nu}g^{\alpha \beta}(x,k)=g^{\mu\nu}(x,k)$. The matrix $\Lambda$ in this appendix is the symmetry generator in which the forms are explained in Sec. \ref{sec: metric}. The dispersion relation essentially a relation between mass and the inner product of momentum, which both are Lorentz invariants.

\end{appendix}



\begin{thebibliography}{99}





\bibitem{Adler:2001vs}
R.~J.~Adler, P.~Chen and D.~I.~Santiago,
Gen. Rel. Grav. \textbf{33}, 2101 
(2001).

\bibitem{Ling:2005bq}
Y.~Ling, B.~Hu and X.~Li,
Phys. Rev. D \textbf{73}, 087702 (2006).

\bibitem{Ling:2005bp}
Y.~Ling, X.~Li and H.~b.~Zhang,
Mod. Phys. Lett. A \textbf{22}, 2749 
(2007).



\bibitem{Galan:2006by}
P.~Galan and G.~A.~Mena Marugan,
Phys. Rev. D \textbf{74}, 044035 (2006).



\bibitem{Myung:2006qr}
Y.~S.~Myung, Y.~W.~Kim and Y.~J.~Park,
Phys. Lett. B \textbf{645}, 393 
(2007).

\bibitem{Liu:2007fk}
C.~Z.~Liu and J.~Y.~Zhu,
Gen. Rel. Grav. \textbf{40}, 1899 
(2008).

\bibitem{Salesi:2009kd}
G.~Salesi and E.~Di Grezia,
Phys. Rev. D \textbf{79}, 104009 (2009).

\bibitem{Gim:2014ira}
Y.~Gim and W.~Kim,
JCAP \textbf{10}, 003 (2014).

\bibitem{Nozari:2015rza}
K.~Nozari, S.~Saghafi and A.~Damavandi Kamali,
Astrophys. Space Sci. \textbf{357}, 
140 (2015).

\bibitem{Mu:2015qna}
B.~Mu, P.~Wang and H.~Yang,
JCAP \textbf{11}, 045 (2015).

\bibitem{Kim:2016qtp}
Y.~W.~Kim, S.~K.~Kim and Y.~J.~Park,
Eur. Phys. J. C \textbf{76}, 
557 (2016).

\bibitem{Lobo:2020oqb}
I.~P.~Lobo and G.~B.~Santos,
Phys. Lett. B \textbf{817}, 136272 (2021).

\bibitem{Song:2025qpo}
J.~Song and C.~Liu,
Int. J. Mod. Phys. A \textbf{40}, 
2550045 (2025).

\bibitem{Rohim:2025gxo}
A.~Rohim, G.~I.~B.~Darman, M.~F.~Fauzi and A.~Sulaksono,
Int. J. Mod. Phys. D \textbf{34}, 
2550047 (2025).

\bibitem{Amelino-Camelia:2011lvm}
G.~Amelino-Camelia, L.~Freidel, J.~Kowalski-Glikman and L.~Smolin,
Phys. Rev. D \textbf{84}, 084010 (2011).

\bibitem{Pfeifer:2021tas}
C.~Pfeifer and J.~J.~Relancio,
Eur. Phys. J. C \textbf{82}, 
150 (2022).



\bibitem{Bonanno:2000ep}
A.~Bonanno and M.~Reuter,
Phys. Rev. D \textbf{62}, 043008 (2000).

\bibitem{Vafa:2005ui}
C.~Vafa,
[arXiv:hep-th/0509212 [hep-th]].

\bibitem{Brennan:2017rbf}
T.~D.~Brennan, F.~Carta and C.~Vafa,
PoS \textbf{TASI2017}, 015 (2017).

\bibitem{Palti:2019pca}
E.~Palti,
Fortsch. Phys. \textbf{67},
1900037 (2019).

\bibitem{Loges:2019jzs}
G.~J.~Loges, T.~Noumi and G.~Shiu,
Phys. Rev. D \textbf{102}, 
046010 (2020).

\bibitem{Amelino-Camelia:2008aez}
G.~Amelino-Camelia,
Living Rev. Rel. \textbf{16}, 5 (2013).
\bibitem{Lukierski:1992dt}
J.~Lukierski, A.~Nowicki and H.~Ruegg,
Phys. Lett. B \textbf{293}, 344 
(1992).

\bibitem{Witten:2024upt}
E.~Witten,
Eur. Phys. J. Plus \textbf{140}, 
430 (2025).

\bibitem{Chen:2014jwq}
P.~Chen, Y.~C.~Ong and D.~h.~Yeom,
Phys. Rept. \textbf{603}, 1 
(2015).

\bibitem{Ishibashi:2021kmf}
A.~Ishibashi, N.~Ohta and D.~Yamaguchi,
Phys. Rev. D \textbf{104}, 
066016 (2021).

\bibitem{Carmona:2019fwf}
J.~M.~Carmona, J.~L.~Cort{\'e}s and J.~J.~Relancio,
Phys. Rev. D \textbf{100}, 
104031 (2019).

\bibitem{Lukierski:2002fd}
J.~Lukierski and A.~Nowicki,
Acta Phys. Polon. B \textbf{33}, 2537 
(2002).

\bibitem{Relancio:2020zok}
J.~J.~Relancio and S.~Liberati,
Phys. Rev. D \textbf{101}, 
064062 (2020).

\bibitem{Carmona:2021gbg}
J.~M.~Carmona, J.~L.~Cort{\'e}s and J.~J.~Relancio,
Universe \textbf{7}, 
99 (2021).

\bibitem{Pawlowski:2018swz}
J.~M.~Pawlowski and D.~Stock,
Phys. Rev. D \textbf{98}, 
(2018).

\bibitem{Born:1938zve}
M.~Born,
Proc. Roy. Soc. Lond. A \textbf{165}, 
291 
(1938).

\bibitem{Yao:2018ceg}
Y.~Yao, M.~S.~Hou and Y.~C.~Ong,
Eur. Phys. J. C \textbf{79}, 
513 (2019).

\bibitem{Binetti:2022xdi}
E.~Binetti, M.~Del Piano, S.~Hohenegger, F.~Pezzella and F.~Sannino,
Phys. Rev. D \textbf{106}, 
046006 (2022).

\bibitem{Cavaglia:2003qk}
M.~Cavaglia, S.~Das and R.~Maartens,
Class. Quant. Grav. \textbf{20}, L205-L212 
(2003).

\bibitem{Ali:2014xqa}
A.~F.~Ali,
Phys. Rev. D \textbf{89},
104040 (2014).

\bibitem{Cavaglia:2004jw}
M.~Cavaglia and S.~Das,
Class. Quant. Grav. \textbf{21}, 4511 
(2004).

\end{thebibliography}
\end{document}